# Prediction of spatial distribution of debris-flow hit probability considering the source-location uncertainty

**Running title**: Probabilistic Debris-Flow Prediction

Kazuki Yamanoi[1,2,]*, Satoru Oishi[2], Kenji Kawaike[1]

[1]Disaster Prevention Research Institute, Kyoto University
[2]Riken Center for Computational Science

*Corresponding author: Kazuki Yamanoi

Mailing address: Yokooji Shimomisu-Higashinokuchi, Fushimi Ward, Kyoto 612-8235, Japan

Phone number: +81-75-611-4397

Fax number: +81-75-611-4397

E-mail: yamanoi.kazuki.6s@kyoto-u.ac.jp

ORCID: 0000-0002-2837-2522

**Acknowledgments**: This work was supported by JSPS KAKENHI Grant Numbers 19K15105 and 21H01436, and FOCUS Establishing Supercomputing Center of Excellence (COE). We would like to thank Editage (www.editage.com) for English language editing.

**Conflicts of interest**: The authors declare no conflicts of interest associated with this manuscript.

# Prediction of Spatial Distribution of Debris-Flow Hit Probability Considering the Source-Location Uncertainty

**Abstract**

Prediction of the extent and probability of debris flow under rainfall conditions can contribute to precautionary activities through risk quantification. To this end, quantifying the debris-flow risk against rainfall involves three components: predicting the debris-flow initiation locations under rainfall conditions, setting appropriate physical parameters related to debris-flow transportation, and evaluating the affected area using numerical simulation. In this study, we developed a logistic regression method that includes rainfall and topographic parameters as explanatory variables to quantify the probability of debris-flow initiation in an actual area with disaster record. Moreover, an objective parameter-set selection was introduced by evaluating the agreement between the simulation results with multiple parameters and the erosion/deposition area determined using the aerial light detection and ranging difference data after debris flow. Finally, by combining these results, we conducted a predictive debris-flow transport simulation using the initiation datasets generated by the logistic model. Therefore, the spatial distribution using the probability of the effects of debris-flow, which can be applied for risk quantification and evacuation optimization, was successfully obtained at 1-m resolution. Furthermore, a real-time hazard probability prediction system could be developed based on the presented simulation cost.

# 1. Introduction

## 1.1 Background

### 1.1.1 Numerical simulation for debris-flow prediction

Debris flows are seriously hazardous to human life, particularly in mountainous and high-precipitation countries. Therefore, debris-flow precaution is essential in flood risk management because debris flow causes damage to human life and property and changes flood risk by depositing sediment in urban river channels. Precautionary measures can be divided into occurrence and affected-area prediction. Occurrence prediction usually employs a statistical method using rainfall observation data to issue warning information, applying the empirical/heuristic method (Polat & Erik, 2020) or sometimes the machine-learning-based method (Di et al., 2019). In contrast, affected-area prediction primarily employs a numerical simulation to predict debris-flow propagation in topographies to estimate the hazard area. However, all hazard areas are not affected by debris flow in an actual situation with occurrence prediction. In addition, the extent of the affected area varies depending on rainfall, which changes the volume of water in the debris flow. Therefore, if both the extent and probability of debris flow depending on rainfall can be predicted, the precautionary activities can be remarkably improved.

The location and properties of the debris sources are essential as the initial requirements for simulating the debris-flow. The source location of the debris flow, usually defined as the scour at the top of the debris-flow trace, is a crucial parameter for determining the motion of the debris flow (Wu & Lan, 2019; 2020). The source location is crucial for determining the affected area; therefore, it is usually set at the actual locations estimated from the field survey or interpretation of aerial photographs to conduct the simulation (Rodríguez-Morata et al., 2019; Schraml et al., 2015). However, physical, empirical, or statistical prediction is required when the simulation is performed without using the pre-existing source-location data. Hence, applying landslide prediction is a possible method to predict source locations. Generally, the predicted location of a landslide is provided as a susceptibility map. Many researchers have already proposed a statistics-based prediction method by analyzing the relationship among morphological, geological, land-cover, and hydrological conditions, as well as landslide locations to estimate the spatial distribution of landslide susceptibility (Reichenbach et al., 2018). Yamanoi et al. (2022) employed cell-by-cell logistic regression, commonly used to evaluate susceptibility, in high resolution to predict the probability of the source locations from terrain conditions. They combined it with a two-dimensional debris-flow simulation considering the propagation and development due to erosion of the bed surface sediment to predict the affected area. However, the method aimed only to estimate the affected area at the same scale of conditions, and thus neglected the inequality of rainfall conditions. In particular, this method does not apply to a wide area with unequal rainfall conditions.

Furthermore, many researchers have already proposed landslide occurrence prediction by inputting spatial and temporal distribution of rainfall (e.g., Guzzetti et al., 2020) and implemented it in real scenarios. Generally, the intensity-duration (ID) method is commonly used to estimate ID thresholds by analyzing past landslide inventories and predicting the occurrence of landslides. However, some studies have proposed a prediction method employing weather radar to consider rainfall distribution on a finer scale than the rainfall-gauge-based method. In Japan, the soil water index (SWI) and 60-min rainfall are used instead of IDs in landslide prediction to issue alerts (Osanai et al., 2010). However, the method predicts only the occurrence, not the quantitative probability nor the specific location of the affected area based on the rainfall conditions. In contrast, the physically-based prediction method, for example, combining the rainfall-infiltration simulation and stability analysis, is already in use (Van den Bout et al., 2021). However, it requires multiple physical parameters, sometimes calibrated with the observation data, leading to difficulties in prediction. Based on the above, adding the SWI, calculated from the rainfall conditions to the cell-by-cell logistic regression method (Yamanoi et al., 2022) will facilitate predictive simulation against rainfall conditions over a comparatively large region without calibration with the rich observation data.

**1.1.2 Spatial resolution of the debris-flow simulation**

Spatial resolution and computational load are trade-off relationships in debris-flow simulations. Thus, these simulations are generally conducted at resolutions of 2–15 m (Abraham et al., in press; Baggio et al., 2021; Bao et al., 2019; Frank et al., 2015; Liu et al., 2021; Rickenmann et al., 2006; Rodríguez-Morata et al., 2019; Schraml et al., 2015; van Asch et al., 2014). Some simulations with a resolution of 0.5–1 m have already been proposed (Armanini et al., 2009; Ouyang et al., 2013); however, they only apply to limited areas far from the catchment scale or multiple debris flows. Moreover, spatial resolution is crucial from the perspective of topographic reproducibility and details of the output information. A high-resolution simulation facilitates the estimation of the damaged area in detail, which can be applied for advanced evacuation planning. Additionally, simulation with a comprehensive digital terrain model helps include narrow channels in residential areas and microtopography in mountainous channels. Furthermore, recent high-resolution elevation difference data observed using airborne light detection and ranging (LiDAR) are currently available, facilitating direct comparison with similar resolution simulation results.

**1.1.3 Parameter estimation**

Debris-flow models typically include physical parameters, such as grain size and internal friction angles. Generally, researchers try to set the values of these parameters based on field surveys; however, accurate values cannot always be set based on limited data because each underground parameter has a spatial distribution. Therefore, trial-and-error parameter calibration is typically performed using other observational data, such as the inundation area. Moreover, the trial-and-error process is not always documented; hence, such a simulation does not always employ optimal parameter sets. In contrast, the inundation area and extensive information on topographic changes can be obtained using airborne LiDAR. The parameter set with less variability and high reproducibility is expected to be estimated by developing a systematic parameter determination using LiDAR differencing information. Additionally, a relationship analysis between the optimized parameter set and geological features might lead to the prediction of the parameter without observation in the future. Therefore, an objective selection of parameter sets within a possible range with the highest reproducibility of LiDAR differences can rationalize the effective debris flow prediction.

### 1.2 Objective

This study aims to develop a methodology for predicting the probability of debris-flow hazard as hit probability in the presence of observed or predicted rainfall distribution. To achieve this, SWI calculated from the rainfall data is added as the variable to the logistic regression model to predict the debris flow sources. In addition, multiple calculations is performed by changing parameter sets within a possible range to estimate the parameter set with the highest reproducibility of LiDAR differences, and the predictive numerical simulation is conducted using this set. These simulations are performed at 1-meter resolution employing a super computer, which allows for considering microtopography and estimating detailed damage extent, and facilitate the direct comparison with the LiDAR observation data. Furthermore, we discuss the feasibility of real-time probability estimation from the viewpoint of computational cost.

## 2. Methodology and material

### 2.1 Target area

The target of this study was the granite area located in the southern Hiroshima Prefecture (Figure 1). The area was affected by a flood in 2018, associated with multiple floods and debris flows caused by heavy rainfall. The number of landslides and debris-flow was approximately 4,000 in Hiroshima Prefecture (Geospatial Information Authority of Japan, n.d.) (Figure 1). In the Sozugawa Basin (Figure 2), multiple landslides, debris flows, and associated floods damaged the downstream area, killed one person, and destroyed 570 houses (Saka Town, n.d.). According to the aerial photograph shown in Figure 2 (a), the downstream residential district was affected by debris flows from multiple sources. The elevation difference observed by LiDAR is shown as Figure 2(b), which is extracted by affected area polygon (Hiroshima University Investigation Team for Torrential Rain Disaster in July 30, 2018).

### 2.2 Debris-flow simulation

For debris flow, we employed a two-dimensional (2D) simulation model presented by Yamanoi et al. (2022) based on Takahashi's model (Takahashi, 2007), which is widely applied for simulating debris-flow movement and inundation caused by debris flow (Jalayer et al., 2018; Uchida et al, 2013) or landslide-dam erosion (Takayama and Imaizumi, 2022). The fundamental equations for this model are the shallow water equations considering multi-phase friction, the equation for transportation of the sediment concentration, and the equation that expresses erosion and deposition. They are described in the conservative vector form as follows:

$$\frac{\partial \boldsymbol{U}}{\partial t} + \frac{\partial \boldsymbol{E}}{\partial x} + \frac{\partial \boldsymbol{F}}{\partial y} = \boldsymbol{S}, \tag{1}$$

$$\boldsymbol{U} = \begin{pmatrix} h \\ uh \\ vh \\ Ch \\ z_b \end{pmatrix}, \boldsymbol{E} = \begin{pmatrix} uh \\ u^2h + \frac{1}{2}gh^2 \\ uvh \\ Cuh \\ 0 \end{pmatrix}, \boldsymbol{F} = \begin{pmatrix} vh \\ uvh \\ v^2h + \frac{1}{2}gh^2 \\ Cvh \\ 0 \end{pmatrix}, \boldsymbol{S} = \begin{pmatrix} i \\ gh(S_{0x} - S_{fx}) \\ gh(S_{0y} - S_{fy}) \\ iC_* \\ -i \end{pmatrix} \tag{2}$$

where $h$ is the water depth, $u$ and $v$ are the depth-averaged velocities in the $x$- and $y$-directions, respectively, $C$ is the concentration of sediment in the fluid, $z_b$ is the land-surface elevation, $g$ is the acceleration due to gravity, $C_*$ is the sediment concentration in the soil at the ground surface, and $S_0x$ and $S_0y$ are the topographical gradients in the x- and y- directions, respectively. For simplicity, the Reynold's stress term in eq. (2) included in Yamanoi et al. (2022) was neglected according to previous research employing similar debris-flow models (Suzuki, 2018; Uchida, 2013).

$S_{fx}$ and $S_{fy}$ are the frictional gradients between the fluid and bed surface in the x- and y-directions, respectively. They were calculated using three different equations according to the sediment concentration to express the three-phase (stony debris, hyper-concentrated, and water) flow. $i$ is the erosion speed estimated by the function $C$ and the equilibrium concentration $C_\infty$, which is also calculated using different equations for the three-phase flow. The equations for the model are listed as follows:

$$S_{fx} = \begin{cases} \dfrac{u\sqrt{u^2 + v^2}d_m^2}{8gh^3\left\{C + (1-C)\dfrac{\rho}{\sigma}\right\}\left\{\left(\dfrac{C_*}{C}\right)^{\frac{1}{3}} - 1\right\}}, & C \geq 0.4C_* \\ \dfrac{u\sqrt{u^2 + v^2}d_m^2}{0.49gh^3}, & 0.01 \leq C < 0.4C_* \\ \dfrac{n_m^2 u\sqrt{u^2+v^2}}{h^{\frac{4}{3}}}, & C < 0.01 \end{cases} \tag{3}$$

$$i = \begin{cases} \delta_e \dfrac{C_\infty - C}{C_* - C_\infty} \dfrac{h\sqrt{u^2+v^2}}{d_m}, & C_\infty \geq C \\ \delta_d \dfrac{C_\infty - C}{C_*}\sqrt{u^2+v^2}, & C_\infty < C \end{cases} \tag{4}$$

$$C_\infty = \begin{cases} 0.9C_*, & \tan\theta_w \geq \tan\phi \\ \dfrac{\rho\tan\theta_w}{(\sigma-\rho)(\tan\phi-\tan\theta_w)}, & \tan\phi > \tan\theta_w \geq 0.138 \\ 6.7\left\{\dfrac{\rho\tan\theta_w}{(\sigma-\rho)(\tan\phi-\tan\theta_w)}\right\}^2, & 0.138 > \tan\theta_w \geq 0.03 \\ \dfrac{\rho(1+5\tan\theta_w)}{\rho-\sigma}\left(1-\alpha_c^2\dfrac{\tau_{*c}}{\tau_*}\right)\left(1-\alpha_c\sqrt{\dfrac{\tau_{*c}}{\tau_*}}\right), & \tan\theta_w \leq 0.03 \wedge \tau_* > \tau_{*c} \\ 0, & \tan\theta_w \leq 0.03 \wedge \tau_* \leq \tau_{*c} \end{cases} \quad (5)$$

$$\tau_{*c} = 0.04 \times 10^{1.72\tan\theta_w}, \quad (6)$$

$$\tau_* = \frac{\rho}{\sigma-\rho}\frac{h\tan\theta_w}{d_m}, \quad (7)$$

$$\alpha_c^2 = \frac{2\left(0.425 - \dfrac{\sigma\tan\theta_w}{\sigma-\rho}\right)}{1 - \dfrac{\sigma\tan\theta_w}{\sigma-\rho}} \quad (8)$$

where $d_m$ is the representative sediment diameter, $n_m$ is the Manning's roughness coefficient, $C_*$ is the sediment density in the river bed, $\delta_e$ and $\delta_d$ are the erosion and deposition coefficient, $\theta_w$ is the water surface gradient, $\sigma$ and $\rho$ are the specific weights of water and sediment, respectively.

Yamanoi et al. (2022) used the MacCormack scheme with diffusion-type artificial viscosity to solve the aforementioned equations in a simulation with 5-m resolution. However, a more satisfactory resolution requires higher robustness, especially at 1-m resolution. Therefore, to increase the robustness, we introduced the total variation diminishing (TVD) scheme (Liang et al., 2007; Ouyang et al., 2013) instead of using the diffusion-type artificial viscosity and used the method proposed by Suzuki et al. (2018) to solve the discontinuity in $S_{fx}$, $S_{fy}$, and $C_\infty$ for each equation.

### 2.3 Landslide prediction

The location and geometry of the landslide are required as an initial condition for the debris-flow simulation. However, the location and geometry of the landslide measured by aerial photography, satellite imagery, or field in situ observations are unavailable for prediction. In this study, we employed the cell-by-cell logistic regression method presented in a previous study (Yamanoi et al., 2022) to predict the probability of debris-flow initiation. The previous study selected terrain slope, water accumulation, tangential curvature, and plan curvature as explanatory variables through regression analysis. However, because the method was applied to a limited area where rainfall was assumed to be uniform, rainfall data were neglected in the regression. In this study, we added the SWI, which represents the precipitation water in the surface soil, to the previous method for application to a relatively large area.

The SWI was obtained as the sum of the water depth in three series of tank models (here, the first tank represents the surface runoff water in the soil layer, the second tank represents the water in the surface infiltration layer, and the third tank represents the underground water) (Ishikawa & Kobatake, 1979) calculated using the following equations.

$$SWI = S_1 + S_2 + S_3,$$
$$\frac{dS_1}{dt} = -\beta_1 S_1 - q_1 + r, \frac{dS_2}{dt} = -\beta_2 S_2 - q_2 + \beta_1 S_1, \frac{dS_1}{dt} = -\beta_3 S_3 - q_3 + \beta_2 S_2 \quad (9)$$

where $S_1$, $S_2$, and $S_3$ are the water depths in the first, second, and third tanks, respectively, and $\beta_1$, $\beta_2$, and $\beta_3$ are the infiltration ratios from first to second tank, second to third tank, and third tank to outside, respectively. $q_1$, $q_2$, and $q_3$ are the runoff discharges from the first to third tanks, which were calculated as follows:

$$q_1 = \begin{cases} \alpha_1(S_1 - L_1) + \alpha_2(S_1 - L_2), & (S_1 > L_2) \\ \alpha_1(S_1 - L_1), & (L_1 < S_1 < L_2) \\ 0, & (S_1 < L_1) \end{cases}$$
$$q_2 = \begin{cases} \alpha_3(S_2 - L_3), & (S_2 > L_3) \\ 0, & (S_2 < L_3) \end{cases}, q_3 = \begin{cases} \alpha_4(S_3 - L_4), & (S_3 > L_4) \\ 0, & (S_3 < L_4) \end{cases}, \quad (10)$$

where $\alpha_1$, $\alpha_2$, $\alpha_3$, and $\alpha_4$ are the runoff ratios for the first to fourth holes in the tank series, and $L_1$ to $L_4$ are the hole heights. The first and second holes were located in the first tank, the third hole in the second tank, and the fourth in the third tank. An important characteristic is that this index represents short- and long-term precipitation. The values of the parameters were applied, as shown in Table 1, which are used in a practical warning system for sediment-related disasters in Japan (Japan Meteorological Agency, n.d.).

We applied a logistic regression model for 10-m resolution cells. Only the granite area was selected as the regression target to reduce the influence of differences in geological features. The debris-flow source locations were interpreted from the aerial photograph; therefore, invisible areas due to cloud coverage were excluded. The target area for the regression is shown in Figure 1. By adding the SWI to the variables used in the previous study (Yamanoi et al., 2022), five types of variables were selected as regression inputs (Figure 3). The maximum SWI was calculated at approximately 250-m resolution employing XRAIN, the real-time precipitation data obtained using the data integration and analysis system (DIAS). The surface slope, accumulation, tangential, and profile curvatures were calculated from the 10-m resolution digital elevation model (DEM) using the r.slope.aspect of the geographic resources analysis support system (GRASS GIS) (Neteler et al., 2012).

The probability of the areas being the source locations of the debris flow at the corresponding cells was calculated using the obtained regression model as follows:

$$p^i = \frac{\exp\{\gamma_0 + \gamma_1 SWI^i_{max} + \gamma_2 \log_{10} A^i_c + \gamma_3 \theta^i + \gamma_4 cr^i_p + \gamma_5 c^i_t\}}{1 + \exp\{\gamma_0 + \gamma_1 SWI^i_{max} + \gamma_2 \log_{10} A^i_c + \gamma_3 \theta^i + \gamma_4 cr^i_p + \gamma_5 c^i_t\}} \quad (11)$$

where $SWI^i_{max}$ is the maximum SWI at cell $i$, $A^i_c$ is the catchment area, $\theta^i$ is the local terrain gradient, and $c^i_p$ and $c^i_t$ are the local curvatures in the plane and tangential directions, respectively. $\gamma_0$ to $\gamma_5$ are the coefficients which were obtained using regression, and their values are shown in Table 2.

The spatial distribution of probability in the target area can be calculated using eq. (11), as shown in Figure 4 (a). Source-location data was generated by providing pseudo-random numbers to the probability values, as shown in Figure 4 (b). This study introduced SWI into the logistic function; hence, the SWI value changed the probability values. As shown in Figure 4 (b), the spatial densities of the generated debris-flow sources were similar to the actual density, implying that the proposed model with SWI successfully estimated the probability without contradictions.

# 3. Numerical simulations for determining appropriate parameter set and prediction

## 3.1 Simulation conditions

The simulation input data are presented in Figure 4. Figure 4 (a) shows the probability of a debris-flow source, obtained using eq. (11), employing the 10-m resolution DEM and precipitation data of 2018. Figure 4 (b) shows the 1-m resolution DEM, with the observed and predicted source locations. In Figure 4 (b), the sources are expressed as points; however, they have an area of 10 m × 10 m because they have been obtained from the 10-m resolution probabilities. The predicted source locations were obtained as shown in Figure 4 (b) by employing a set of pseudo-random numbers and the probability distribution shown in Figure 4 (a). Only three cases are shown in Figure 4 (b); however, 100 cases of the simulation were conducted in this study.

The debris-flow initiation at multiple source locations was simulated as follows: At $t = 0$ [s], the debris-flow mass was set at the source locations (10 m × 10 m) as a fluid consisting of water and sediment. Owing to the characteristics of this simulation, a high sediment concentration causes rapid deposition at the source location. Hence, we increased the initial water volume in the initial debris-flow mass and added the subsequent water flow after the debris-flow initiation to avoid the rapid sediment deposition. Under this condition, the initial sediment concentration was assumed to be half of the concentration in the bed material $C_*$. As the water volume increased, the initial depth of debris-flow increased to 1.5 times the soil depth $D_e$. The subsequent water flow was set using the water discharge per unit area $Q$ [m/s] and duration $T_d$. Further, $r_c$ is the transition rate of the finer particles into the fluid material based on the phase shift concept (Uchida et al., 2013). The relationship among the $r_c$, specific weight of the fluid material $\rho$, and sediment density in the river bed

$C_*$ are described as,

$$\rho = 1 + r_c\sigma \quad (12)$$
$$C_* = C_{*0}(1 - r_c). \quad (13)$$

As the $r_c$ increased, the specific weight of the fluid material in the ground $\rho$ increased, whereas $C_*$ decreased. $C_{*0}$ is the sediment density in the river bed without considering the phase shift.

The parameters related to the debris-flow transport such as mean grain size $d_m$, internal friction angle $\phi$, and transition rate $r_c$, which are appeared in the Equations (3), (4), (5), (7), (12), and (13), vary for different sites, particularly with respect to geological features. Erodible soil depth $D_e$ also varies for different sites depending on the geological features and weathering process, which defines the minimum value of $z_b$ in Equation (2) as $z_{bini} - D_e$, where $z_{bini}$ is the initial value of $z_b$. Moreover, $Q$ and $T_d$ are subject to rainfall conditions and infiltration processes; however, the infiltration characteristics also differ depending on the geological features. Generally, each appropriate parameter can be set using direct field surveys and the trial-and-error method. However, we used only the elevation difference obtained from airborne LiDAR observations to determine the appropriate parameter set for the following reasons. First, parameter setting by field survey is challenging when practically applying the proposed method for real-time prediction. Second, airborne LiDAR data can be readily collected. In Japan, the government conducts large-scale LiDAR observations after large-scale sediment-related disasters. Using the accumulated data, acceptable parameter settings may be established based on geological features. Therefore, we conducted 240 cases of simulations, using combinations of the parameter candidates shown in Table 3, employing the supercomputer FUGAKU (Riken Center for Computational Science, Japan). After the simulation, we compared all simulation results with the observed elevation difference to determine the case with the highest degree of agreement.

### 3.2 Evaluation of the simulation results for parameter selection

The calculated topographic changes at $T = 3600$ [s] are shown in Figures 5 (a) and (b) for the two representative cases among the 240 cases. This study employed the confusion matrix, which is the evaluation matrix for the classified variables, to qualify the degree of agreement when comparing the simulation results with observed elevation difference. We classified cells $i$ using the final topographic change $\Delta z_b$ in the simulation and observation into three classes using eq. (12).

$$Class(i) = \begin{cases} \text{Erosion} & \text{if } \Delta z_b < 0, \\ \text{Not affected} & \text{if } \Delta z_b = 0, \\ \text{Deposition} & \text{if } \Delta z_b > 0. \end{cases} \quad (12)$$

Figures 5 (c) and (d) show the spatial distribution of the three-category classification comparing the simulation and observation for the cases shown in Figures 5 (a) and (b), respectively. To compare the simulation with observation classes, the number of cells accounting for true positive (TP), true negative (TN), false positive (FP), and false negative

(FN) were calculated for "Erosion," "Not affected," and "Deposition" as shown in Tables 4, 5, and 6, respectively. Furthermore, *precision*, *recall*, and *F*1 score (the harmonic mean of precision and recall), are defined as follows:

$$precision = \frac{TP}{TP + FP}, recall = \frac{TP}{TP + FN},$$
$$F1 = \frac{2 \times precision \times recall}{precision + recall} = \frac{2TP}{2TP + FP + FN}, \quad (13)$$

A higher *F*1 score indicates a higher *precision* and *recall*, thus indicating a high correspondence between the simulation and observation. In this study, the average of the three *F*1 values, $F1_{ave}$, for "Erosion," "Not affected," and "Deposition" was used to compare all the cases quantitatively. Figure 6 shows the *F*1 scores in all the different simulation cases using the six representative parameters. In this study, we selected the parameter set with the highest *F*1 score, shown in italics in Table 3, and whose simulated topographic change is shown in Figure 5 (a).

### 3.3. Predictive numerical simulation

By changing the distribution of source locations created by the set of pseudo-random numbers, partially shown in Figure 4 (b), we conducted 100 Monte Carlo debris-flow simulations. This set of simulations can be considered as predictive simulation because it inputs the initiation locations predicted from rainfall data. Topographic change maps at $T = 3600$ [s] were obtained from the simulation using the supercomputer, as shown in Figure 7. We obtained a similar pattern of deposition in the downstream area by accumulating numerous debris flows in the catchment topography, as shown in the previous study (Yamanoi et al., 2022). The average topographic change and relative standard deviation values are shown in Figures 8 (a) and (b), respectively. As the elevation difference increased in the valley bottom or enclosed area of the river, the relative standard deviation decreased around the main river streams. The relative frequency of the debris-flow effect, which is defined as the area of $|\Delta z_b| > 0$, is shown in Figure 8 (c). This figure indicates the probability of the debris-flow effects associated with the distributed rainfall that determines the location of the debris-flow sources, assuming that other factors, such as the physical parameters shown in Table 3, are less dominant.

## 4. Discussion

A 1-m resolution simulation, which facilitated the consideration of urban channel structures, was successfully conducted at a catchment scale, which is a larger domain compared to the previous studies (Armanini et al., 2009; Ouyang et al., 2013). The spatial distribution of the debris-flow affected probability was also estimated at a 1-m resolution as shown in Figure 8 (c). This study did not include the effect of buildings; however, such fine-resolution information is generally used for performing evacuation in a microscale urban area. In addition, this study introduced the SWI into the source-location prediction model.

Therefore, the occurrence probability of the source location was developed considering the distributed rainfall pattern. The spatial density of the source location did not contradict the actual density. Therefore, the model with SWI was used to estimate the debris-flow sources based on the rainfall data.

The maximum SWI was calculated from a specific rainfall event; however, it can be replaced with real-time SWI based on rainfall observations or near-future SWI based on short-term rainfall forecasting. Therefore, if the time consumed for all simulation cases is sufficiently short, the real-time probability can be obtained theoretically. The actual time required to compute the one-hour debris flow was 1286 [s] on average with a standard deviation of 440 [s], using 384 nodes (i.e., 1536 MPI parallel) on the FUGAKU, supercomputer. The FUGAKU has 158,976 nodes; thus, 400 cases can be simulated simultaneously. Hence, assuming the use of a 60-min rainfall forecast and neglecting the time for pre-processing (e.g., rainfall index calculation and source-location prediction) and post-processing (e.g., calculation of the affected frequency), the method predicts the future from 60–120 min in approximately 30 min. Thus, this method can be used for the development of a real-time hazard probability prediction system. The real-time hazard probability facilitates optimized evacuation, that is, selecting the shelter or evacuation route with the lowest hazard probability. Furthermore, this information can reduce unnecessary evacuations with a high casualty risk.

This study uses the averaged $F1$ score of the deposition and erosion area as an evaluation function of the debris-flow simulation to estimate the optimal parameter set. The current simple method changes the parameter set in the possible range and determines the parameter set with the highest agreement; however, an effective metaheuristic method for parameter optimization can be implemented by expanding upon the current method. Furthermore, additional observational data, such as water level during rainfall obtained by the simple gauges, whose number has been increased by the Japanese government, can also be applied for parameter estimation.

## 5. Conclusions

In this study, we developed the prediction and generation methodologies of the debris-flow initiation locations from rainfall and topographical data to achieve the predictive debris-flow simulation under rainfall conditions. We also developed an appropriate parameter-set selection method by quantifying the degree of agreement between the simulated and observed topographic change. By using this set, we performed the predictive debris-flow simulations in 1-m resolution by inputting the generated initiation location patterns. Finally, by calculating the relative frequency of the debris-flow reaches, the maps of the probabilistic debris-flow affected areas were proposed.

However, this study has some limitations. The proposed relative frequency is not a strictly predicted hazard probability due to the following reasons. First, this study used LiDAR difference data in parameter-set selection, which can be obtained only after the debris-flow event. However, for the prediction, a set of appropriate parameters must be provided in advance, which depends on the geological features. Moreover, it is difficult to assume that only one parameter set can be determined for each geological feature. Thus, the uncertainties of the multiple parameters must be considered under several geological features. The uncertainty of each parameter can be expressed as a probability density function using calculation methods, such as Bayesian estimation. In this case, we can design a similar Monte Carlo simulation method by changing the parameters according to probability in addition to the probability for initiation locations. Second, the duration and scale of the subsequent water flow were included in the input parameters, which might be strongly affected by rainfall conditions. To consider such effects, not only the location and number of debris flow but also subsequent flow should be predicted based on the rainfall conditions. The accurate prediction of the subsequent hydrograph may be difficult because of the limitations of the observation data; however, the simulation of underground water transport has the potential to solve these problems. Future research can apply the abovementioned strategies, in addition to those used in this study, to estimate a more strict and appropriate hazard probability.

## Data availability statement

The datasets generated and analyzed during the current study are available from the corresponding author on reasonable request.

**Tables**

Table 1. Parameters for the three series of tank models.

|  | First tank | Second tank | Third tank |
|---|---|---|---|
| Height of hole (mm) | $L_1 = 15$, $L_2 = 60$ | $L_3 = 15$ | $L_4 = 15$ |
| Runoff ratio (1/h) | $\alpha_1 = 0.1$, $\alpha_2 = 0.15$ | $\alpha_3 = 0.05$ | $\alpha_4 = 0.01$ |
| Infiltration ratio (1/h) | $\beta_1 = 0.12$ | $\beta_2 = 0.05$ | $\beta_3 = 0.01$ |

$\alpha_1$, $\alpha_2$, $\alpha_3$, and $\alpha_4$ are the runoff ratios for the first to fourth holes in the three tanks in series; $\beta_1$, $\beta_2$, and $\beta_3$ are the infiltration ratios to each tank; and $L_1$ to $L_4$ are the hole heights.

Table 2. Results of the logistic regression.

|  | Value | Std. error | z-value | p-value |
|---|---|---|---|---|
| $\gamma_0$ | -18.5350 | 0.3805 | -48.7111 | $< 10^{-15}$ |
| $\gamma_1$ | 0.0417 | 0.0017 | 25.1712 | $< 10^{-15}$ |
| $\gamma_2$ | -0.7764 | 0.0604 | -12.8494 | $< 10^{-15}$ |
| $\gamma_3$ | 0.0467 | 0.0020 | 23.2909 | $< 10^{-15}$ |
| $\gamma_4$ | 10.2806 | 1.9080 | 5.3882 | 7.1169e-08 |
| $\gamma_5$ | -37.8506 | 1.5203 | -24.8971 | $< 10^{-15}$ |

Table 3. Simulation parameter candidates. Italicized values show the parameter set selected using the mean $F1$ score.

| Parameter [unit] | Description | Candidate values |
|---|---|---|
| $d_m$ [m] | Mean diameter | *0.02*, 0.05, 0.1 |
| $D_e$ [m] | Erodible depth | 1, *2* |
| $\phi$ [deg.] | Internal friction angle | 35, *25* |
| $r_c$ | Transition rate | 0, *0.1*, 0.2 |
| $Q_{add}$ [m/s] | Subsequent water discharge | *0.1*, 0.2, 1, 10 |
| $T_{add}$ [s] | Duration of subsequent water | 1, 10, *100*, 500, 1000 |

Table 4. Confusion matrix for "Erosion."

| | | Prediction | | |
|---|---|---|---|---|
| | | Erosion | Not affected | Deposition |
| | Erosion | TP | FN | FN |
| Observation | Not affected | FP | TN | TN |
| | Deposition | FP | TN | TN |

TP, true positive; TN, true negative; FP, false positive; FN false negative.

Table 5. Confusion matrix for "Not affected."

|  |  | Prediction | | |
|---|---|---|---|---|
|  |  | Erosion | Not affected | Deposition |
| Observation | Erosion | TN | FP | TN |
|  | Not affected | FN | TP | FN |
|  | Deposition | TN | FP | TN |

TP, true positive; TN, true negative; FP, false positive; FN false negative.

Table 6. Confusion matrix for "Deposition."

|  |  | Prediction | | |
|---|---|---|---|---|
|  |  | Erosion | Not affected | Deposition |
| Observation | Erosion | TN | TN | FP |
|  | Not affected | TN | TN | FP |
|  | Deposition | FN | FN | TP |

TP, true positive; TN, true negative; FP, false positive; FN false negative.

**Figures**

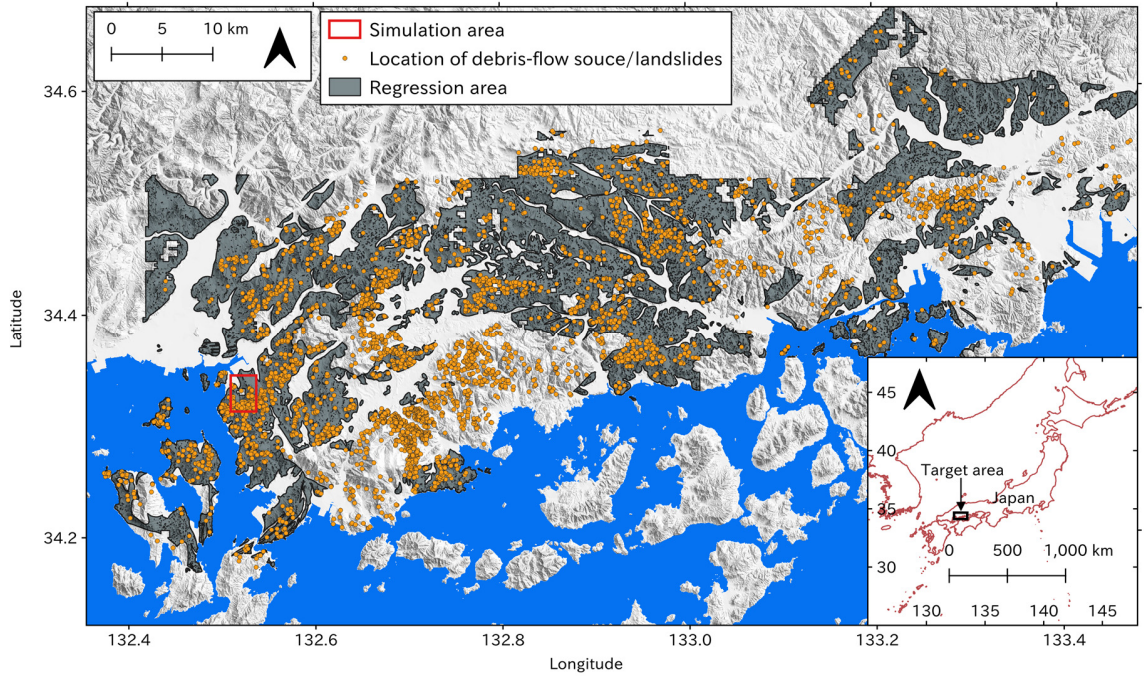

Figure 1. An overview of the target area of this study located in the southern part of Hiroshima Prefecture, Japan. The yellow points represent the location of the debris-flow source or landslides extracted from the aerial photograph by the Geospatial Information Authority of Japan. The regression area is defined as the area where the geology is classified as granite, excluding the cloud cover area in the aerial photograph.

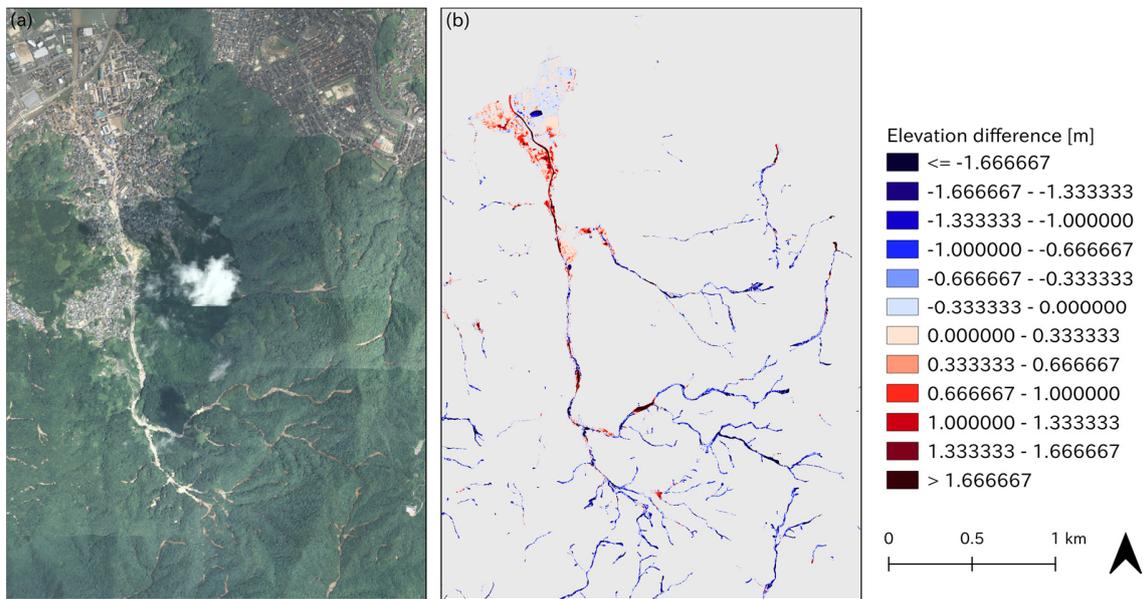

Figure 2. An overview of the simulation area. (a) The aerial photograph taken after the 2018 flood event by Geospatial Information Authority of Japan. (b) The elevation difference observed by airborne light detection and ranging (LiDAR) of the affected area polygon (Hiroshima University Investigation Team for Torrential Rain Disaster in July 30, 2018).

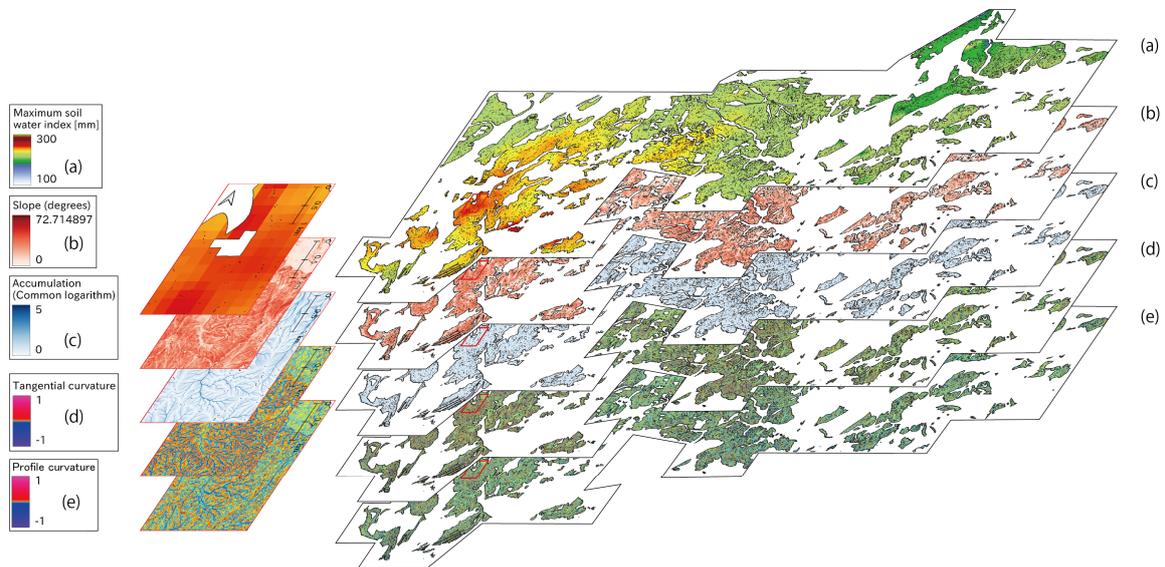

Figure 3. Regression inputs. (a) Maximum soil water index in the 2018 flood event calculated from XRAIN, the radar-based rainfall data. (b) Terrain slopes. (c) Water accumulation are shown as the common logarithms. (d) Tangential curvature. (e) Profile curvature. (b), (c), (d), and (e) are calculated using the digital elevation model with 10-m resolution provided by the Geospatial Information Authority of Japan and using r.slope.aspect and r.watershed in Geographic Resources Analysis Support System (GRASS).

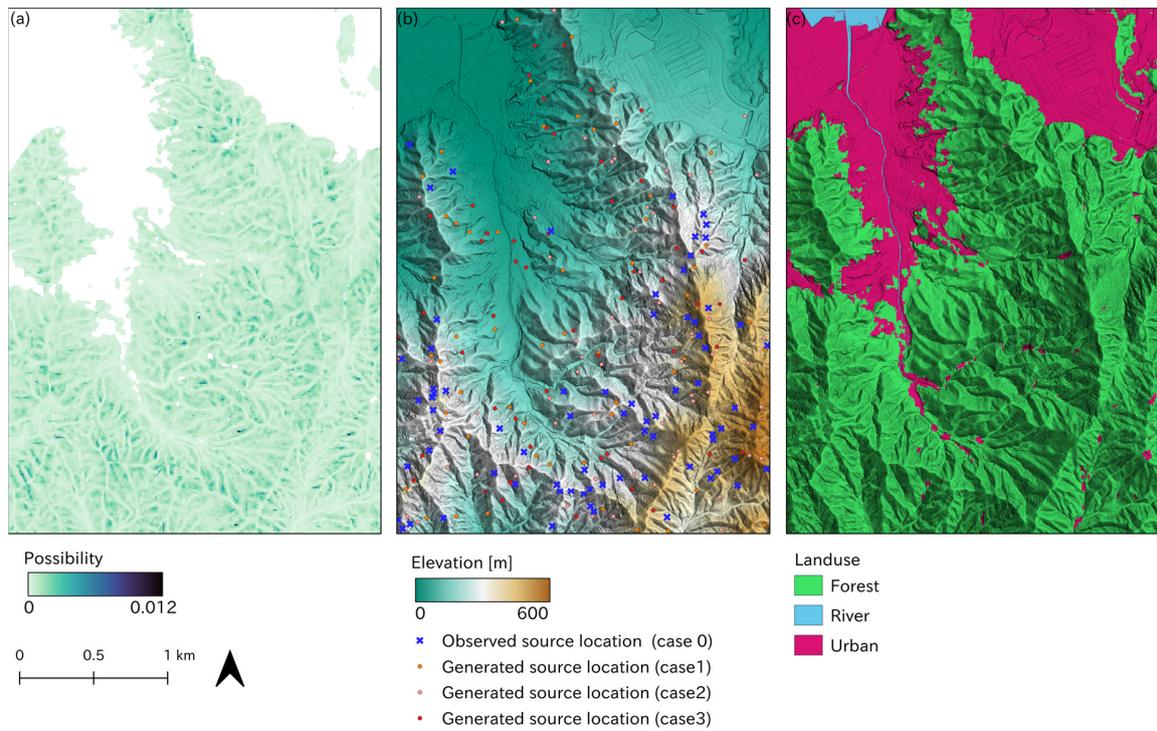

Figure 4. Simulation inputs. (a) Estimated debris-flow source probability based on the 2018 heavy rain conditions. (b) Elevation with 1-m resolution and debris-flow source locations. (c) Land-use conditions, which are classified into three categories.

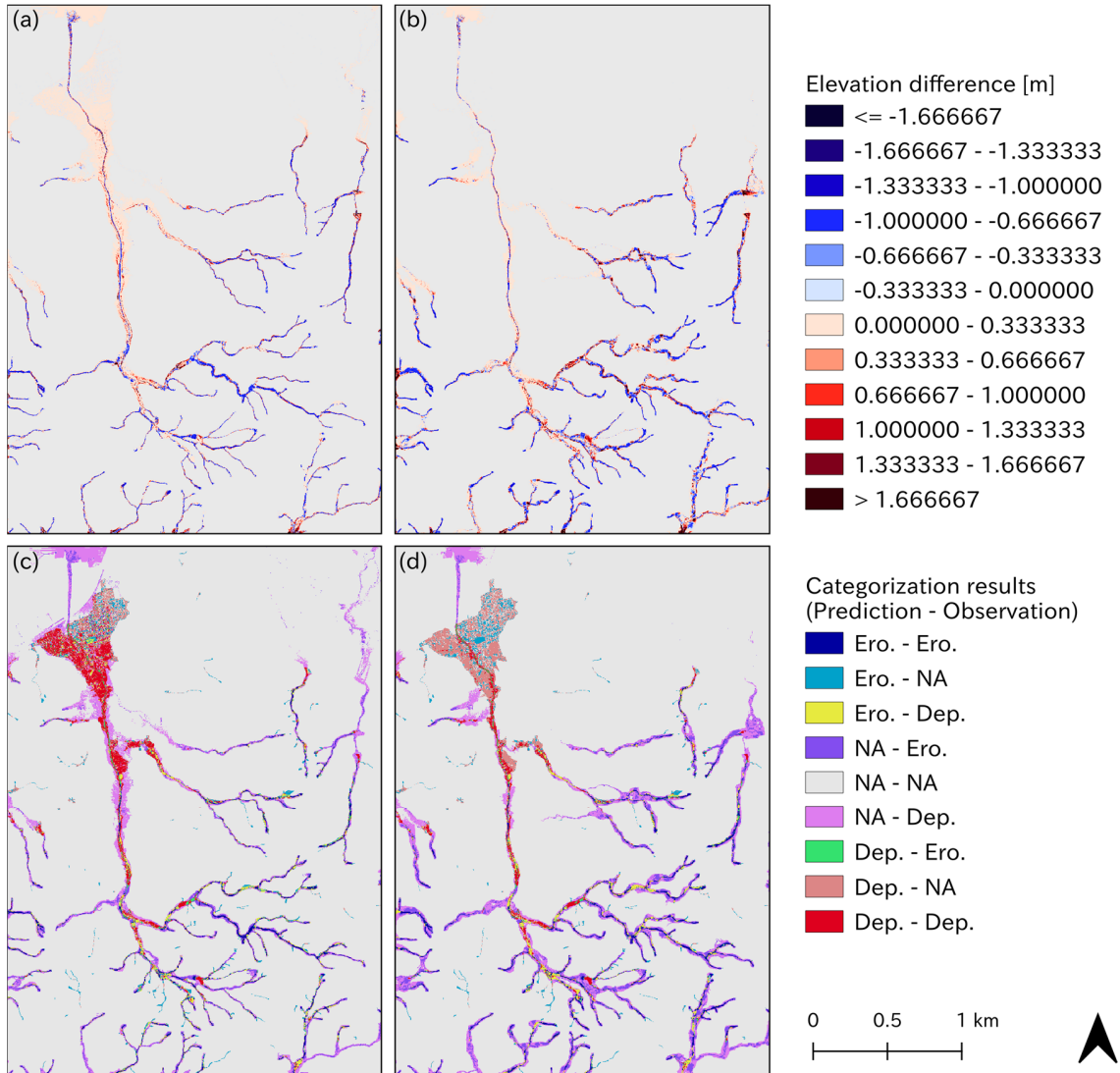

Figure 5. Simulation results. The results of final topographic change $\Delta z_b$ (a) at prediction time $T = 3600$ [s] with mean diameter $d_m = 0.02$ [m], erodible soil depth $D_e = 1.0$ [m], $\phi = 25$ [deg.], transition rate $r_c = 0.1$, $Q_{add} = 0.1$ [m/s], and $T_{add} = 500$ [s] and (b) at $T = 3600$ [s] with $d_m = 0.02$ [m], $D_e = 1.0$ [m], $\phi = 35$ [deg.], $r_c = 0.2$, $Q_{add} = 1$ [m/s], and $T_{add} = 10$ [s]. (c) and (d) are the classified maps for (a) and (b) combined with the observation, respectively.

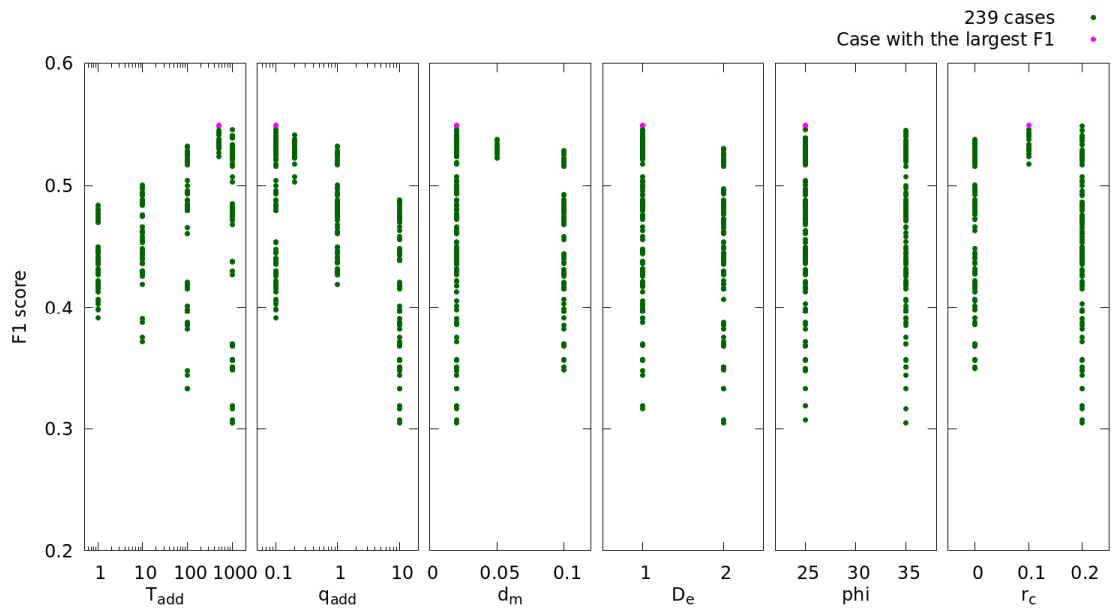

Figure 6. Relationship between $F1$ scores and parameters in all the simulation cases. The pink colored dot represents the parameter sets with the highest $F1$ scores.

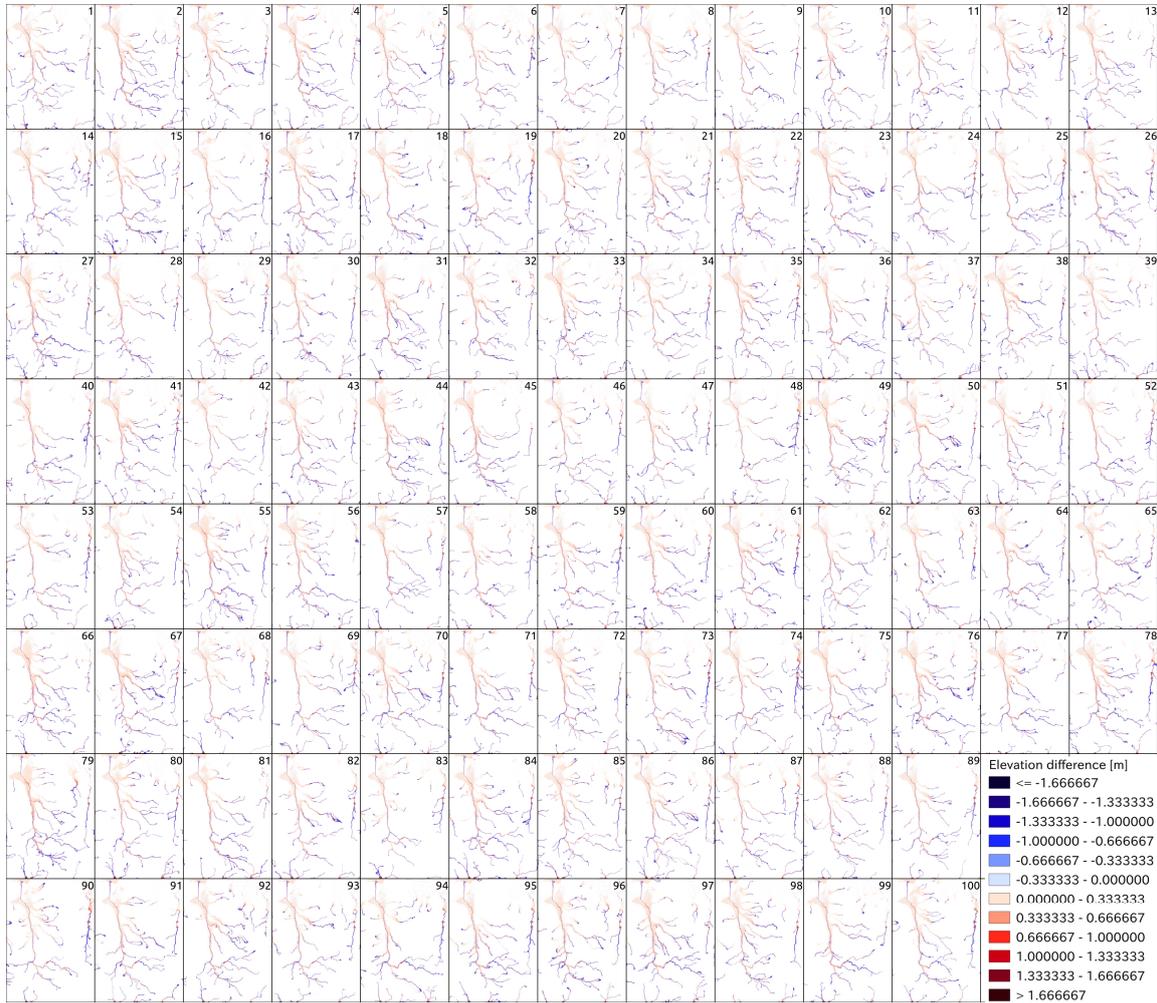
Figure 7. Simulation results of topographic change for 100 cases obtained by changing the source-location distribution.

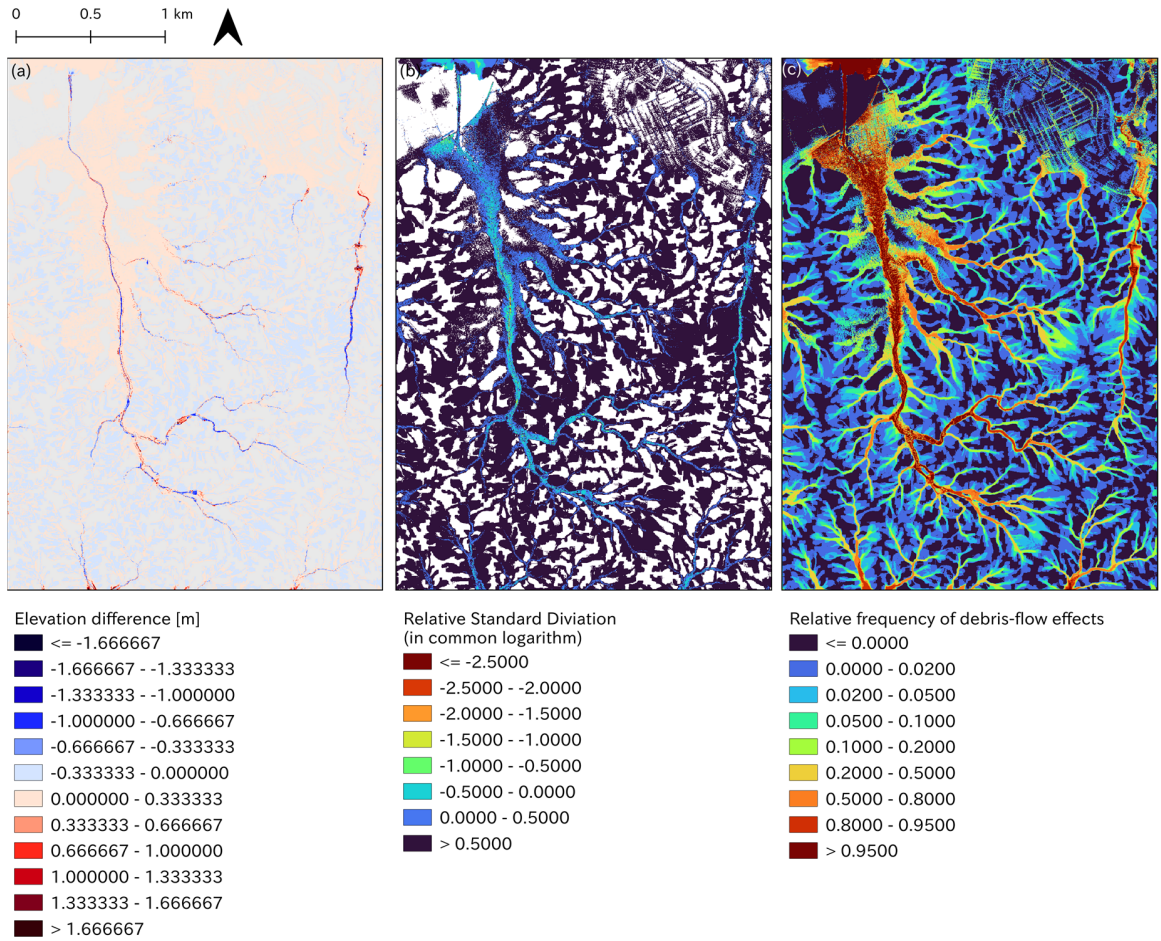

Figure 8. Ensemble simulation results obtained by changing the source-location distribution. (a) Average elevation difference. (b) Relative standard deviation shown as common logarithms. (c) Relative frequency corresponding to the debris-flow affected area in 100 cases.